\documentstyle[11pt,aaspp4]{article}
\tighten

\journalid{337}{}
\articleid{11}{14}

\def\spose#1{\hbox to 0pt{#1\hss}}
\def\lta{\mathrel{\spose{\lower 3pt\hbox{$\mathchar"218$}}
     \raise 2.0pt\hbox{$\mathchar"13C$}}}
\def\gta{\mathrel{\spose{\lower 3pt\hbox{$\mathchar"218$}}
     \raise 2.0pt\hbox{$\mathchar"13E$}}}
\def\n{\noindent}
\def\cl{\centerline}

\def\be{\begin{equation}}
\def\ee{\end{equation}}
\def\msun{M_{\odot}}
\def\rsun{R_{\odot}}

\def\mdot{\dot M}

\def\mpy{M_{\odot} {\rm yr}^{-1}}
\def\aa#1{{Acta Ast} {#1}}

\begin{document}

\title {Circumbinary Disks and Cataclysmic Variable Evolution} 
\author{ H. C. Spruit\altaffilmark{1} and Ronald E. Taam\altaffilmark{2}}

\n \altaffilmark{1}{Max-Planck-Institut f\"ur Astrophysik, Postfach 1317, D-85741
Garching, Germany}

\n \altaffilmark{2}{Department of Physics \& Astronomy, Northwestern 
University, Evanston, IL 60208}

\begin{abstract}
The influence of a circumbinary (CB) disk on the evolution of cataclysmic 
variable (CV) binary systems is investigated. We show that CB mass surface 
densities sufficient to influence the evolution rate are plausibly provided 
by the outflows observed in CVs, if the net effect of these winds is to  
deliver $10^{-4}$--$10^{-3}$ of the mass transfer rate to the CB disk. The torque
exerted by the CB disk provides a positive feedback between mass transfer rate
and CB disk mass 
which can lead to mass transfer rates of $\sim 10^{-8} -10^{-7} \mpy$.  This
mechanism may be responsible for causing the range of variation of mass transfer
rates in CV's. In 
particular, it may explain rates inferred for the novalike variables and 
the supersoft X-ray binary systems observed near the upper edge of the 
period gap ($P \sim 3 - 4$ hr), as well as the spread in mass transfer rates 
above and below the period gap. Consquences and the possible observability of
such disks are discussed.
\end{abstract}

\keywords {binaries: close --- stars: evolution ---- stars: cataclysmic
variables}

\section{INTRODUCTION}

Angular momentum losses play an essential role in the origin and 
evolution of close binary systems 
containing compact objects. An excellent example is the class of systems 
known as cataclysmic variable (CV) binaries in which a low mass star 
transfers mass to its more massive white dwarf companion. Significant
orbital angular momentum must have been lost with ejected mass during 
the formation process to transform a long period progenitor system   
into a CV via the common envelope evolutionary phase (see reviews by 
Iben \& Livio 1993 and Taam \& Sandquist 2000). On the other hand,  
angular momentum loss without significant mass loss is thought to 
occur during the secular evolution of CV's as a result of gravitational  
radiation (Paczy\'nski 1967; Faulkner 1971) or magnetic braking  
(Verbunt \& Zwaan 1981) processes. 

The need for a process such as magnetic braking for the long 
term evolution of CV's stems from the fact that the inferred mass transfer
rates can exceed that driven by gravitational radiation by more than 
an order of magnitude. Although the inclusion of such processes in 
evolutionary models has been modestly successful, detailed comparisons
of observations with theory lead to some difficulties. Specifically, 
it is found that the mass transfer rates primarily 
depend on the time scale of the 
angular momentum loss and the mass of the donor with little dependence 
on the white dwarf companion.  This leads to a strong correlation  
between mass transfer rate and orbital period.  
The inferred mass transfer rates do not follow such 
a tight correlation (Patterson 1984; Warner 1987); there is about an 
order of magnitude spread at a given orbital period.  This discrepancy
has been further aggravated with the discovery of supersoft
X-ray sources among the short period cataclysmic variable population. 
In particular, three sources have been detected with orbital periods
near the upper edge of the period gap, i.e.,  
J0439.8-6809 at 3.37 hr (Schmidtke \& Cowley 1996), J0537.7-7304 at
3 hr (Orio et al. 1997), and 1E 0035.4-7230 at 4.13 hr (Crampton et 
al. 1997).  Such sources are believed to be systems for which the 
white dwarf component accretes at rates $\sim 1 - 4 \times 10^{-7} 
\mpy$, sufficient for steady hydrogen burning to take place on its 
surface (van den Heuvel et al. 1992).  

In order to explain the large spread in mass transfer rates for 
a given orbital period, suggestions involving intermittent cycles 
produced by nova explosions (Shara et al. 1986) or by irradiation or 
mass loss effects (King et al. 1996) have been invoked.  However, 
the recent lack of detection of the mass losing component in novalike 
variables in the infrared spectral wavelength region by Dhillon et al. 
(2000) suggests that the high mass transfer
rates in these systems ($\mdot \sim 10^{-8} \mpy$) are secular rather
than cyclic. This would cause them to be expanded and cool compared with 
main sequence stars in the same range of orbital periods. A range in  
evolutionary states for the mass losing star has also been suggested for 
the spread in mass transfer rates by Pylyser \& Savonije (1988), but 
the dearth of systems inside the period gap is hard to reconcile with this 
idea.  
 
As a possible resolution to 
these problems in our understanding of the evolution of CVs,
we suggest that a circumbinary (CB) disk can effectively drain   
orbital angular momentum from the system, promoting mass transfer at rates
above those calculated using the magnetic braking or gravitational 
radiation process. This is in contrast to the self-excited winds picture
proposed by King \& van Teeseling 1998).  In such a 
situation the orbital angular momentum can be removed from the binary by 
tidal torques (see Lin \& Papaloizou 1979; Eggleton \& Pringle 1985; 
Pringle 1991). We present here a model 
incorporating such a disk into the secular evolution 
of CV systems. In the next section, we present the assumptions of the 
model and describe the properties of such disks and their critical role
for the secular evolution. 

The implications of these results are discussed in section 3. There we also
address a number of observational puzzles that may be related to the presence 
of circumbinary disks. In addition to the large spread in mass transfer rates, 
this includes the low luminosity of the secondaries in novalike systems. We also
touch upon the origin of the period gap in CVs, a problem that arguably is
aggravated rather than solved by a CB disk picture.

\section{THEORETICAL PICTURE}

We assume that a CB disk has formed in a CV system. A fossil CB disk may 
have been established at an early stage as a result of its formation    
following the common envelope phase.  Alternatively, a CB disk may be  
formed as a result of mass outflow from the white dwarf or the accretion disk. 

Studies of UV and optical resonance lines in the bright CV systems (novalike
variables, dwarf novae in outburst, and supersoft sources) have yielded evidence of
outflowing material (see Warner 1995).  We now make the assumption that a (small)
fraction of this matter settles into the orbital plane outside the binary system, 
forming a CB disk. Evidence that makes the presence of such material plausible is
the relatively small width of the single-peaked emission lines in the so-called SW
Sex stars (Thorstensen et al. 1991, Hoard 1998, Hellier 2000) and supersoft sources
(e.g. RX J0019.8+2156, Deufel et al. 1999). The half-width of these peaks
corresponds to velocities of the order 500-800 km/s, i.e. not more than about twice
the orbital speed. Though the lines typically also show extended wings indicating
higher outflow speeds characteristic of the inner disk regions, this shows that a
large part of the outflow is ejected with low velocities. 
Theoretically this could, for example, happen when a part of the outflow takes
place in the form of a slow wind near the orbital plane, generated in the outer
regions of the disk as a magnetically driven flow (e.g. Spruit \& Cao 1994). We
will leave the details of this hypothetical process open for the moment, except for
noting that some of the orbital phase-dependent anomalies seen in the SW Sex stars
may be related to material accumulating just outside the orbit. In the model by
Horne (1999), for example, these anomalies are explained in terms of a magnetic
`propeller effect' acting on (a part of) the accretion stream 
(but see Hellier 2000 for an alternative view). 

A possible analogy of the process may occur in Be-stars, where 
the observed cool disks are thought to be formed by compression of the  
radiation-driven stellar wind into the equatorial plane (Bjorkman and Cassinelli
1993, Bjorkman and Wood, 1995), perhaps involving a clumpy flow (Howk et al. 
2000) or from a magnetically driven outflow (Balona \& Kaye 1999; 
Smith \& Robinson 1999).

In contrast to circumstellar accretion disks, CB disks cannot be treated as 
steady since the action of viscosity can in principle make it spread 
indefinitely. A continuous input of mass and angular momentum at its inner 
boundary as in the model proposed here leads to a continuous increase in 
its surface mass density. This is in contrast to the case of accretion disks, 
where the surface density is closely related to the instantaneous mass flux. In
CB disks, the surface density is typically much higher, for the same mass flux. The 
properties and evolution of thin viscous Keplerian CB disks has been investigated
by Pringle (1991) in the context of cool protostellar disks and by Lee, Saio, \&
Osaki (1991) as related to disks around Be stars.  

Starting with an initially empty CB disk, the mass transfer driven by a 
standard magnetic angular momentum loss process from the secondary feeds 
mass into the CB disk. The gravitational interaction of the disk with the 
binary transfers angular momentum from the binary to the disk at a rate 
proportional to the surface density in the inner regions of the disk. The 
gradual buildup of a disk mass implies a growing torque which eventually 
starts dominating over the magnetic wind torque. Our assumption that 
the mass fed into the CB disk is a certain fraction of the binary mass transfer 
rate, then implies a positive feedback: the larger the disk mass, the 
larger the torque and the larger the mass transfer rate. In this way, we 
may expect an eventual runaway to large mass transfer rates, as a result of 
an initially innocuous stellar wind torque. 

The evolution of a CB disk, coupled to the evolution of its mass-providing 
binary is complicated by the detailed physics of binary evolution as well  
as that of the disk itself. These are beyond the scope of the present work. 
Instead, the next section presents a simple model which illustrates the 
basic mechanism, as well as its likely sensitivity to the detailed physics 
that is not included.

\subsection{A Simple Model}

\label{model}
A torque $T$ exerted on a binary with a Roche-lobe filling 
secondary star causes mass transfer from the secondary to the primary 
accretion disk at a rate $\dot M_2<0$ given by
\be
(5/3-2q+\zeta)\dot M_2/M_2=-2T/J. \label{mdot}
\ee
where $J$ is the orbital angular momentum, q is the mass ratio of the
system, and $\zeta={\rm d}\ln R_2/{\rm d}
\ln M_2$ the mass-radius exponent of the secondary (e.g. Frank, King \& 
Raine 1985). We envisage the torque $T$ as made up of two components: a standard
stellar wind torque $\dot J_{\rm w}$ and a torque $T_{\rm e}$ due to the
gravitational interaction of the binary with its CB disk.

In order to arrive at a model system that can be analysed by analytical  
means we make some simplifications. The most important assumption, as 
discussed above, is that of the mass transferred from the secondary to   
the primary, a fraction $\delta$ is fed into the CB disk:
\be \dot M_{\rm e}=-\delta\dot M_2.\label{dotme}\ee
In the following we keep this fraction fixed in time, and
assume it enters the CB disk with a specific angular momentum equal to that of the
inner edge of the disk. We assume $\delta$ to be small, so that eq (\ref{mdot}),
which assumes conservative evolution, is still approximately valid. In addition, we
ignore the secular evolution of the binary parameters, $M_2$,
and $\zeta$ (this will be relaxed in section \S \ref{secular}). Finally, a simple
form is used for the viscosity of the CB disk, specified below.

The viscous evolution of the CB disk is governed by
\be
\partial_t \Sigma={3\over r}\partial_r[r^{1/2} \partial_r({r^{1/2}}\nu
\Sigma)], \label{visdis}
\ee
where $\Sigma$ is the surface mass density and $\nu$ the viscosity, 
which at this point can be an arbitrary function of $r$ and $\Sigma$. 

The viscous torque in the disk is
\be T= -2\pi r^2\nu\Sigma r\partial_r\Omega, \label{torq}\ee
where $\Omega$ is the local orbital frequency $\Omega=(GM/r^3)^{1/2}$, and $M$ 
the mass of the binary system. Approximating the interaction of the binary 
with the disk as taking place locally at the inner edge of the disk, the 
torque $T_{\rm e}$ exerted by the disk on the binary is then
\be
T_{\rm e}= 3\pi (r_{\rm i}/a)^{1/2}\Omega_0 a^2  
\nu_{\rm i}\Sigma_{\rm i}, \label{etor}
\ee
where $\Omega_0$ and $a$ are respectively the binary orbital frequency 
and separation and the subscript $_{\rm i}$ denotes the inner edge of the CB
disk. The rate of change of the orbital angular momentum $J=\mu(GMa)^{1/2}$ (where
$\mu$ is the reduced mass $M_1M_2/M$) of the binary due to the combined effects of
magnetic braking and the CB disk is thus
\be 
-\dot J/J=T/J=\dot J_{\rm w}/J+T_{\rm e}/J=t^{-1}_{\rm w}+
3 \pi (r_{\rm i}/a)^{1/2} {1+q\over M_2} \nu_{\rm i} \Sigma_{\rm i},
\label{ttorque}
\ee
where $t_{\rm w}$ is the angular momentum loss time scale due to 
magnetic braking. 

Suppose that an CB disk is initially absent, and forms by the mass fed   
into it. The initial evolution is then dominated by the first term in 
(\ref{ttorque}). As the disk mass increases, the second term eventually 
becomes important. We first consider the case when the stellar wind torque 
dominates, then the case when the CB disk torque dominates. At the end we 
calculate, numerically, the case with both torques present.

Suppose the viscosity $\nu$ is a function of $r$ only (see below for 
a more general case):
\be \nu=\nu_{\rm i}(r/r_{\rm i})^n, \ee
where $r_{\rm i}$ is the inner edge of the disk. Eq. (\ref{visdis}) can 
then be written as
\be \partial_\tau y=x^{2n-2}\partial _{xx} y, \label{diff}\ee
where
\be 
x=(r/r_i)^{1/2}, \qquad y=3\pi x\nu\Sigma, \qquad \tau=t/t_{\rm vi},
\ee
and $t_{\rm vi}$ is the viscous time scale at the inner edge:
\be t_{\rm vi}={4 r_{\rm i}^2\over 3\nu_{\rm i}}.\label{vii} \ee
The mass flux at any point in the disk is then
\be \dot M=-\partial_x y.\ee
The evolution of the disk is governed by (\ref{diff}), with 
$y\rightarrow 0$ for $x\rightarrow\infty$ and the mass input rate 
providing a boundary condition at $r_{\rm i}$:
\be (\partial_x y)_{\rm i}=-\dot M_{\rm e}. \ee

As an example, consider the case $n=1$, i.e. the viscosity increases linearly with
distance. The $x-$dependent coefficient in (\ref{diff}) is then a constant. By a
simple shift $x\rightarrow x-1$, the boundary condition at $x=1$ can then be moved
to x=0. The problem consisting of equation plus boundary conditions then has no
intrinsic length scale in it. If, as assumed, the initial disk is empty, 
the problem has no intrinsic time scale either. Under these conditions, the
solution  has a self-similar form:
\be y=\tau^\mu f(\xi),\ee
where the {\it similarity variable} $\xi=x\tau^\lambda$ (see Zel'dovich and Raizer, 1986 for a
detailed analysis and physical interpretation of self-similar solutions of
diffusion equations).
In the present case, one finds by substitution into the equation that
$\lambda=-1/2$, and by substitution into the boundary condition that $\mu=1/2$, and
the problem reduces to finding the solution to an ODE for $f$:
\be f''+{1\over 2}\xi f'-{1\over 2}f=0,\qquad{\rm with} f'(0)=-1.\ee

The solution which vanishes at infinity is a parabolic cylinder function $U(a,\xi)$
(as defined in Abramowitz \& Stegun, 1964). In terms of the original physical
variables, the solution can be written as
\be
\Sigma(r,t)=({t\over t_{\rm vi}})^{1/2}{\dot M_{\rm e}\over 3\pi 
\nu_{\rm i}} ({r\over r_{\rm
i}})^{-3/2} A e^{-\xi^2/8} U(a,\xi), \qquad {\rm with}\quad \xi = (x - 1)/ 
\tau^{1/2},
\ee
where the parameter $a=1/\sqrt 5$ and $A$ is a numerical constant close to unity. 

The asymptotic behavior of $\Sigma$ is as $e^{-\xi^2/4}$. The disk thus has
has an effective edge at $\xi\sim 1$, which corresponds to a physical distance $r_{\rm d}$
\be
[(r_{\rm d}/r_{\rm i})^{1/2}-1]\approx (t/t_{\rm vi})^{1/2}.
\ee
For large times, the surface density thus increases with time as 
$t^{1/2}$, and decreases with distance as $r^{-3/2}$, with a cutoff at 
$r_{\rm d}$. This cutoff distance increases linearly with time. Evaluating now the
torque 
on the binary due to the CB disk for this case, we find that 
(cf. \ref{ttorque})
\be 
t_{\rm e}^{-1}\equiv {T_{\rm e}\over J}=
-\left({r_{\rm i}\over a}\right)^{1/2}{1+q\over M_2} B(t/t_{\rm
vi})^{1/2} \dot M_{\rm e},
\ee
where $B=AU(a,0)\approx 1.3$. With $r_{\rm i}/a\approx 1.7$ (see below),
and $q=0.5$, this yields, with
(\ref{dotme}), 
\be 
t^{-1}_{\rm e}\approx 3\delta (t/t_{\rm vi})^{1/2} \vert\dot M_2\vert/M_2. 
\ee

The assumption that the dominant torque is the stellar wind torque breaks 
down when $t_{\rm e}$ becomes of the same order as the wind angular 
momentum loss time scale $t_{\rm w}$. From (\ref{ttorque}) we find that 
this occurs at a time $t_0$,
\be
t_0\approx 0.1 t_{\rm vi}/\delta^2.\label{t0}
\ee
To get an order of magnitude estimate for this time scale, assume a disk with inner
edge radius $\sim 2\times 10^{11}$ cm around a 1M$_\odot$ binary. If the
temperature of this disk is about 1000K (the sort of number that the calculations
below will yield), the viscous time scale is $2/\alpha$ yr for a standard
$\alpha$-viscosity prescription. For $\alpha=0.01$, the order of magnitude of the
viscosity found in simulations of magnetic turbulence, we have $t_{\rm vi}=200$ yr.
Taking $\delta=10^{-3}$ then gives $t_0\approx 2 \times 10^7$ yr. 

After this time, the evolution accelerates due to the CB disk torque. To 
find the time dependence in this phase, we can neglect the contribution of 
the stellar wind torque. The CB disk mass input $\dot M_{\rm e}$ now increases 
in proportion to the surface density. Using (\ref{mdot}) and (\ref{ttorque}), 
the inner boundary condition can be expressed as 
\be
(\partial_x y)_{\rm i}=-k y_{\rm i},\label{larget}
\ee
where
\be k=\delta {2(r_{\rm i}/a)^{1/2}(1+q)\over 5/3-2q+\zeta}. \ee
The problem is now mathematically homogenous in time, and the behavior at 
large time is found as the most unstable normal mode of eqs. (\ref{diff}) 
and (\ref{larget}), with $n=1$. There is only one unstable mode:
\be y\sim e^{k^2\tau-kx}.\ee
With the parameters previously used, with $\zeta\approx 0.8$ and $k\approx 
3\delta$, the growth time of the instability is of the order $t_{\rm inst}
\approx 0.1 t_{\rm vi}/\delta^2$, which is of the same order as $t_0$.

Hence, for a viscosity independent of surface density and linearly
proportional to distance, the surface density of the CB disk
increases as $t^{1/2}$, up to a time $t_0$. After this, the growth becomes
exponential on a time scale of the same order as $t_0$. The whole time evolution is
thus governed by a {\it single time scale} $t_0$. Apart from a factor of 
order unity this is the viscous time scale at the inner edge of the
CB disk, divided by the square of the fraction of the mass that ends 
up in the CB disk. [For more general viscosity prescriptions, the dependence on
$\delta$ is different, see below]. The value of $t_0$ is in the `interesting' range
for CV evolution of $10^7$--$10^9$ yr if $\delta\approx 10^{-4}$--$10^{-3}$. This 
is an agreeably small number that would not put too strong a constraint on 
the mechanism feeding the CB disk.

The transition between the two stages in the evolution of the disk, when both
torques are important, cannot be obtained with simple analytic means. Full
solutions of the problem were obtained numerically using an implicit (Crank
Nicholson) scheme. The result, shown in Fig. 1, illustrates the analytic trends. In
this calculation, the binary parameters $q$, $M_2$, $\zeta$ are again artificially
kept fixed during the disk evolution, as in the above  analytic estimates.

\subsection{Disk Physics}

The evolution time scale turns out to be sensitive to changes in the 
viscosity in the disk; not just at its inner edge where the torques act, 
but also to its value at larger distances. This is because, unlike the case 
of an accretion disk where the local surface density is simply proportional 
to the mass flow, the surface density in a CB disk depends on its entire 
evolution. 

To determine the sensitivity of the results of the previous section on the 
effects of different assumptions about the viscosity, consider now a more general
case where the viscosity scales as powers of both surface mass density and radius,
$\nu  = \nu_i(\Sigma/\Sigma_i)^m (r/r_i)^n$. In this case, the boundary at $x=1$
can not be transformed to the origin any more. As a result, the solutions are not
strictly self-similar as they were in the case $n=1$. They are still {\it
asymptotically} self-similar, however, for $t\rightarrow\infty$. For large times,
the length scale $\Sigma_{\rm i}/ \partial_r\Sigma_{\rm i}$ becomes large compared
to $r$. The inner boundary condition at $x=1$ can then be replaced by one at $r=0$
without causing much error so that the explicit length scale disappears. The
(constant) mass input rate does not introduce a time scale either, and the problem
again has self similar solutions. Analysis of eq. (\ref{visdis}) as in the previous
section then yields
\be 
\Sigma_{\rm i} \approx \left({t\over t_{\rm vi}}\right)^p 
{\dot M_{\rm e} \over 3\pi 
\nu_{\rm i}} \label{tdep}
\ee
with  $p={{m+1}\over 2(2m+2-n)}$. In addition, we find that $\Sigma \sim r^q$ with
$q = -{{n+1/2}\over m+1}$, for $r$ well inside the outer edge of the disk. 
With the time dependence (\ref{tdep}), the CB evolution time scale $t_0$ becomes,
for $q=0.5$, $r_{\rm i}/a=1.7$, $\zeta=0.5$:
\be t_0/t_{\rm vi}\approx (5\delta)^{-1/p}. \label{pdenp} \ee
Since the time scales relevant for CV evolution are long compared with the 
viscous time at the inner edge of the CB disk, the time scale $t_0$ tends to 
become rather sensitive to the viscosity indices $(n,m)$. This is also 
illustrated in Fig.~1. The solid line shows a case in which the viscosity 
is the same as in the previous example, out to $r/r_i=50$, but changes to 
an $r$-dependence with $n=1.2$ outside this radius. As long as the disk has 
not spread to $r/r_i=50$, the evolution is as before, but after this it 
speeds up compared with the previous case. Define, for this case, 
the disk evolution time scale as the time when the CB torque on the binary equals
the wind torque. The seemingly innocuous change of the outer disk viscosity then
reduces the evolution time scale by a factor 10.

The relevant values of $m$ and $n$ and hence the time index $p$ in (\ref{pdenp})
depend upon the physical conditions in the CB disk, and these conditions are not
dissimilar to the conditions found in circumstellar disks around
pre-main sequence stars (e.g., Bell \& Lin 1994).  This follows from 
the fact that the masses of CV's and T Tauri type stars are similar 
($\sim 1 - 2 \msun$). In addition, the 
inner radii of these disks are also comparable in view of the orbital  
separation ($\sim 10^{11}$ cm) of the short period CV's. Since the ratio of 
the inner disk radii to orbital separation ($r_{in} \sim 1.7 a$) 
are relatively insensitive to mass ratio (Artymowicz \& Lubow 1994),  
inner disk radii are a few $\rsun$ which are typical for disks surrounding 
T Tauri type stars. 

To obtain an estimate of the physical conditions in the CB disk, let us 
assume that it has built up to a surface density such that the torque exerted 
on the binary causes it to evolve on a time scale $\tau_{\dot J}\sim 10^8$ yr. 
Using (\ref{etor}) one finds that, for a binary system of $1.4 \msun$ and 
orbital period of 4.5 hrs, 
\be
\Sigma \sim 100 T^{-1}_3(\alpha \tau_{\dot J8})^{-1}{\rm~~g cm^{-2}},\label{bls}
\ee 
where $\alpha$ is the standard disk viscosity parameter, $T=10^3T_3$ the 
temperature and $\tau_{\dot J8}=\tau_{\dot J}/10^8$ yr. With $\alpha$ given, 
the viscous dissipation rate in the disk is known, and one-zone models for 
the local disk structure can be used to determine the temperature, in the same
way as in accretion disks. From the results of Bell \& Lin (1994) we find 
mid-plane temperatures around 1500 K, relatively insensitively to column
densities, in the range $\Sigma\sim 10^3$ to $10^4$ g cm$^{-2}$, corresponding to
disk 
masses $\gta 10^{-7} - 10^{-6} \msun$. Under these conditions, 
the disk structure is determined by the opacity due to grains. For $\alpha 
\sim 10^{-2}$, eq. (\ref{bls}) then indicates column densities of $10^4$ g 
cm$^{-2}$.  With the same one-zone models for the local disk structure, we 
then find that $p$ ranges from 0.25 to 0.67. The higher value applies where 
opacity is due to the effect of metal grains, at temperatures of $\sim 
1300$ K.  

This estimate indicates that variations of the index $p$ by several tenths 
are to be expected as the disk mass builds up and the dominant sources of 
opacity change. When such a change takes place, the sensitivity of the 
evolution time scale to $p$ would cause the evolution to significantly speed up 
or slow down, resulting in either a more dramatic runaway, or a stabilization of 
the torque and mass transfer.  It is clear that a more realistic picture of 
the evolution of the CB disk requires a detailed study of the physical 
conditions in these disks, in particular the opacities.

A further complication to the physical conditions in the CB disk could be 
irradiation by the central source. This is especially important in those 
regions of the disk where the opacity and hence the optical depths are 
low (in the grain evaporation regime). In this case, for strong  
irradiation, we find that $p \sim 0.4$, which is bracketed 
by the above range. 

\section{IMPLICATIONS}

As described in \S II, the influence of the CB disk on the binary system is
determined by the magnitude and time dependence of the torque. 
Although we have obtained some estimates for the time dependence of the torque, 
based on the local physics of 
the disk, the full time dependent description of the disk coupled with the  
binary system is beyond the scope of the present investigation. 
In the following we qualitatively outline the evolutionary possibilities 
that the existence of such a disk provides and discuss its implications 
for various anomalies observed in CV systems.

\subsection{Secular Evolution of Cataclysmic Variables}

\label{secular}

Since a CB disk is intrinsically nonstationary, the torque it exerts at any
moment depends on the  history of mass fed into it. The CB contribution will
exceed the stellar wind torque (assumed here to act in the same manner as in
previous studies) at some time $t_0$, when a certain amount of mass has
accumulated in the disk. If the fraction $\delta=-\dot M_{\rm e}/\dot M$ of the
mass fed into the disk (see \S \ref{model}) is small, $t_0$ is large and the
evolution of the binary proceeds as in the standard magnetic braking scenario
(e.g., Pylyser \& Savonije 1988). In this case the orbital period and the mass
transfer rate decrease with time due to the decreasing mass of the secondary
(e.g. Kolb and Ritter 1990).

On the other hand if $\delta$ is large enough, $t_0$ can become less than the
magnetic braking time scale. Our results show that the evolution then becomes
unstable after $t_0$, with large mass transfer rates resulting. As the mass
transfer accelerates, the evolution of the secondary starts deviating from its
course under magnetic braking alone. The secondary will get further out of
thermal equilibrium. As previous experiments with enhanced rates of angular
momentum have shown (Kolb and Baraffe 1999), this tends to reverse the decrease
of orbital period with time: the system `bounces' at a certain minimum orbital
period. 
The secondary mass is already rather small at this point in time. Since the CB disk
is still present, angular momentum continues to be lost even when the secondary
already has lost most of its mass. A dramatic effect of mass-transfer-fed CB disk
is thus that it can lead to {\it dissolution of the secondary within a finite
time}. 

To demonstrate this, we have computed the evolution of the binary in a slightly
more realistic model. The change of radius of the secondary star is now modeled
by assuming homology, in the same way as in Spruit and Ritter (1983). This model
includes, in an analytic approximation, the thermal effects of nuclear burning
and radiative energy loss at the surface of the star, and gives a fair
approximation for the period gap in the standard disrupted magnetic braking
model. The magnetic braking torque $\dot J_{\rm w}$ is assumed to operate such
that the angular momentum time scale $J/\dot J_{\rm w}$ is constant in time. The
viscous evolution of the CB disk is treated in the same way as in section \S
\ref{model}. Figure 2 compares the evolution with and without a CB disk. In the
absence of a CB disk, the secondary mass, and the mass transfer rate, decline
exponentially with time (solid line). When the feedback due to a CB disk is
included (dotted line), the secondary dissolves after a finite time. 
The mass transfer rate remains limited, a few $10^{-8}M_\odot$/yr in this example,
even in the final phases. The dissolution would therefore not be a very dramatic
event.

The homology model still has artifacts that prevent quantitative application. For
example, the homology relations fix the adiabatic mass-radius exponent at $-1/3$.
Real stars have larger values, which also depend on time as mass is stripped from
the star. The evolution of real stars will therefore differ significantly, though
we expect that they will still dissolve on a finite time scale if feedback by a
CB disk is effective.

The inclusion of a CB disk into the secular evolution 
can lead to the following three possibilities, in order of increasing importance
of the disk torque: (i) evolution similar to that with magnetic braking, but at
slightly enhanced levels of mass transfer, (ii) initial evolution as in (i), but
followed by accelerated mass transfer as the disk torque increases, and (iii)
accelerated evolution to high mass transfer rates soon after the onset of Roche
lobe overflow. Without detailed modeling of these possibilities, we can already
use the qualitative results obtained above to interpret some observational
puzzles.

\subsubsection{The Large Spread of Mass Transfer Rates}

The observed large range in mass transfer rates of systems, at a given orbital
period, is somewhat puzzling in the magnetic braking scenario.  
In the CB interpretation, 
variations are expected to result from the different epochs at which
they formed (i.e., evolution time relative to $t_0$) and from the 
different parameters of the binary system at the onset of mass 
transfer (e.g., orbital period and mass of the donor star)  
as well as from possible variations in the CB feeding parameter $\delta$. 

An increase in the secular (long-term) mass transfer rate could explain 
systems near the upper edge of the period gap ($P \sim 3 - 6$ hr) 
provided that the mass transfer rates
lie in the range $1.5 \times 10^{-9} - 10^{-8} \mpy$ (Baraffe \& Kolb 2000). 
The observational result that the secondary stars can deviate noticeably from
field main sequence stars provides evidence for the secular origin of this
spread (Beuermann et al. 1998). 

Recently, Baraffe \& Kolb (2000) have found that nuclearly evolved 
secondary components are required to reproduce the observed late spectral 
types in systems with orbital periods greater than 6 hr. To prevent these 
systems from entering into the period gap, they suggest that the mass 
transfer rate must increase as the orbital period decreases. An increase (with
decreasing orbital period) in mass transfer rate above that provided by magnetic
braking alone may also be key to understanding the lack of dwarf novae in the 
period range of 3 - 4 hr (see Shafter 1992). 

Shorter period systems, below the period gap, may also require enhanced angular
momentum loss. If magnetic braking is absent below the period gap, as in the
standard version of the disrupted magnetic braking model for the period gap (see
Rappaport, Verbunt, \& Joss 1983; Spruit \& Ritter 1983) the variety of outburst
behaviors of the nonmagnetic CV's is hard to understand. In the framework of a
thermal-tidal instability model for example, the extreme SU UMa stars known as ER
UMa systems require mass transfer rates significantly greater (by a factor of
$\gta 5$) than the rate given by gravitational radiation (Osaki 1996).  

The essential process in the CB disk scenario outlined here is the gravitational
interaction between the orbiting mass donor and the CB disk. The angular momentum
loss rate by this process depends on both the mass of the CB disk and its
previous history. Even with a fixed recipe for the mass input rate into the CB
disk such as we have adopted here, the different ages of the systems will cause a
spread in mass transfer rates. This may be sufficient to account for the rather
diverse mass transfer rates observed in CVs, and could be the underlying
parameter determining the membership of a CV to one of the subclasses: the U Gem,
Z Cam, novalikes and supersoft sources above the period gap, and the SU UMa, ER
UMa and permanent superhumpers below the period gap.

\subsubsection{The distribution of CV's below the period gap}
Additional evidence in favor of enhanced mass transfer rates for CV's 
below the period gap is the difference between the calculated and 
observed distribution of systems near and above the period minimum at $\sim$ 80
minutes. 

Population synthesis investigations (e.g. de Kool, 1992; Kolb 1998; Kolb \& 
Baraffe 1999) predict a strong frequency increase of systems towards shorter
periods, if the angular momentum loss is of the order of the gravitational
radiation loss rate. This conflicts
significantly with observations, which show no evidence of this accumulation.
Enhanced angular momentum loss is an obvious solution to this problem. Kolb \&
Baraffe (1999) suggest that rates $\sim 4$ times greater than the gravitational
radiation rate would bring the predicted period minimum into agreement with
observations. A mild acceleration of mass transfer by the CB disk mechanism could
plausibly provide this. 

An interesting side effect of the mechanism is that it can also remove systems
{\it permanently} from the CV population as discussed by Patterson (1998) since, 
as discussed above, it can
dissolve the secondary in a finite time, of the order of the mass transfer time
scale. In the standard scenario with angular momentum loss by gravitational
radiation, or a residual magnetic braking, this does not happen because mass
transfer slows down so much that dissolution of the secondary takes
longer than a Hubble time (cf. Figure 2).

\subsubsection{The High Mass Transfer Rates in Supersoft Sources 
and Nova-like Variables}
 
The mass transfer rates inferred for supersoft sources and novalike 
variables are the highest ($\gta 10^{-8} - 10^{-7} \mpy$) among the known 
short period CV's. The recently discovered supersoft sources  
J0537.7-7304, J0439.8-6809, and 1E 0035.4-7230 in the period range of 3 to 
4.13 hr accentuate the difference between observation and theory.  
The difficulty in forming these systems stems from the requirement 
of simultaneously reproducing the high mass transfer rates and the short 
orbital period of the systems. Thermally unstable mass transfer scenarios 
have been discussed by King et al. (2000), and found to be limited in producing,
especially, the shortest period system of this class, J0537.7-7304. In the CB disk
interpretation, these systems would be in the CB disk dominated phase, with
accelerated mass transfer to high levels. In contrast to the thermally unstable  
mass transfer interpretation where high mass secondaries ($\gta 1 
\msun$) and initial orbital periods greater than about 10 hr are 
required, these systems could have evolved from systems with shorter orbital
periods, with secondary masses $\lta 0.3 \msun$. Thus we interpret them as 
high mass transfer rate extensions of the novalike variables.  

\subsection{Other Implications}
\subsubsection{The Low Luminosity of the Secondaries in Nova-likes}

In the standard evolutionary scenario for CV's, the novalike
variables would have had the same average angular momentum loss 
as the dwarf novae, with mass transfer rates at $\sim 10^{-9} \mpy$.  
Their current higher mass transfer rates could be accommodated provided 
that they would be compensated by lower rates at other times (see \S 1). 
At this average rate, the secondary star is only weakly out of thermal 
equilibrium and hence should appear much like an ordinary main 
sequence star.  Its mass and expected luminosity can then be derived
from the orbital period. With this predicted luminosity the secondaries in 
several nearby novalike systems should be easily detectable
in the infrared. With only one exception none have been found (Dhillon et 
al. 2000), in several cases with strong upper limits. In our interpretation, 
this shows that these secondaries are actually significantly expanded and
much less massive than main sequence stars at the same orbital period. This 
requires large {\it average} mass transfer rates over time scales of the 
order $10^7$ yr. In our CB disk mechanism, this is a natural outcome 
since it predicts that the current high mass transfer rates  
are also representative of the past $10^7$ yr.

This interpretation may not apply to all novalikes, however. The detection
by Dhillon et al. (2000) of the secondary in the longer-period novalike 
variable RW Tri (with the longer orbital period of 5.57 hr) suggests that it has
not undergone significant mass loss. In the CB interpretation, this could be a 
system that relatively recently has switched to a phase of high mass transfer.

\subsubsection{Magnetic CVs}

A special group of CV's, from the CB perspective, are the magnetic CVs (AM Her
stars or polars). Since no disk is present in these systems, the CB-feeding disk
outflow postulated here would also be absent. One would thus predict that magnetic
CVs would not have the large mass transfer rates of novalikes, and a smaller spread
in transfer rates. Observational evidence indicates that the dispersion in mass
transfer rates for magnetic CV's is indeed smaller than in non-magnetic CVs
(Patterson 2000). The secular evolution of these systems is then expected to be
similar to that described by the standard magnetic braking scenarios. 

\subsection{Complications and unsolved problems}

\subsubsection{The effect of nova outbursts}

CVs are all believed to have periodic nova-outbursts, in which a significant amount
of mass ($\sim 10^{-4}M_\odot$) is ejected at large velocity. The question thus
arises whether the relatively tenous and weakly bound CB disks can survive nova
outbursts. For the CB
parameters estimated in the above, i.e. inner radii $\sim 2\times 10^{11}$ cm,
temperature around 1000K, and $\alpha$-viscosity $\sim 0.01$, the aspect ratio of
the CB disk is $H/r\sim 5\times 10^{-3}$. This is also the solid angle subtended by
the disk as seen fom the WD. If the mass surface density is $10^3-10^4$ as required
to cause a substantial CB torque under these conditions, the CB disk mass is about
$10^{-7} \msun$. For an isotropic nova explosion, the mass intercepted by the CB
disk is then of the same order as its own mass. In view of the significant
uncertainty in our estimate of the masses of CB disks (notwithstanding the possible
presence of a fossil disk), it is at this point not
clear whether or not they would survive nova outbursts. More detailed study is
needed to establish the likely CB disk masses. We note, however, that this concern does
not apply to the supersoft sources, which are steadily burning their accreting
mass, hence will not produce nova outbursts.

\subsubsection{The period gap}

Whereas CB disks may solve some of the observational puzzles discussed above, it is
only fair to say that they do not much improve our understanding of the period gap
in CVs (the low frequency of systems in the 2--3 hr orbital period range). As
discussed in \S 3.1.1, a large spread in mass transfer rates results in a loss of
coherence in the CV evolutionary tracks, which has a significant effect on the
period gap.  

On the one hand, those systems in which the mass transfer is slightly enhanced, but
not accelerated (see \S 3.1) would still form a period gap, under the standard
disrupted braking hypothesis. These would be in the dwarf nova subclass, and the
transfer rates from ZAMS donors would have to lie in the range $1 - 2 \times
10^{-9} \mpy$ at the upper edge of the gap (Baraffe \& Kolb 2000). On the other
hand, systems with CB-induced accelerated mass transfer as described in \S 2
would not have evolved in this way. Instead, the ones with the largest transfer
rates would have experienced a period bounce at longer orbital periods, and never
evolve into a short-period systems. This effect is reminiscent of Whyte and Eggleton's
(1980) proposal for forming the upper edge of the period gap, and might
contribute to its existence. If there is a range of CB disk masses however, as we
have suggested, the result would be a significant blurring of the upper edge as 
a result of systems detaching at different orbital periods (see McDermott \& 
Taam 1989). 
Also, the CB hypothesis does not explain the preference of novalike systems for the
3-4 h period range, just above the gap.

\subsection{Observability of CB disks}

The most important test of the CB hypothesis is, of course, a direct detection  
of the circumbinary material. This is not likely to be very easy, however. 

The luminosity $L_{\rm e}$ of the CB disk follows directly from the torque 
it exerts on the binary. If the torque is $T_{\rm e}$ [see eq.~(\ref{torq})],
$L_{\rm e}=\Omega_0 T_{\rm e}$ where $\Omega_0$ is the orbital frequency. If the
CB torque dominates, this can be related to the accretion luminosity $L_{\rm a}$:
\be L_{\rm e}/L_{\rm a}\approx R_1/a\,\delta \approx 10^{-2} \delta 
\sim 10^{-5},\ee
where $R_1\approx 10^9$ cm is the radius of the primary star. 

Whether a CB disk of such low luminosity can be detected against the bright
background of the accretion disk depends very much on its spectral energy
distribution. 
At an accretion rate of $10^{-8} \mpy$, $L_{\rm a}\sim 10^{35}$ erg s$^{-1}$, 
$L_{\rm e}\sim 10^{30}\delta_{-3}$ erg s$^{-1}$. Since both the surface density 
and shear rate drop
with radius, most of $L_{\rm e}$ is emitted near the inner edge. For a binary of
period 3.5 hr, the effective emitting area is of the order $A\sim \pi r_{\rm
i}^2\sim 6\times 10^{22}$ cm$^2$. The effective temperature is then $\lta 1000$ K,
suggesting that CB disks should perhaps be detectable in the L band, as a
continuum contribution due to dust emission. 

The outer part of the accretion disk would also contribute in this wavelength
region, however. With a radius $r_{\rm d}\approx 0.3a\approx 3\times 10^{10}$ cm
and luminosity $L_{\rm d}\approx GM\dot M_{\rm a}/r_{\rm d}\approx 3\times
10^{33}$ erg s$^{-1}$, its surface temperature would be around 15000 K. In the 
infrared around
$4\mu$ the accretion disk would then be about as bright as the CB disk. This may
make it rather hard to detect the CB disk by its broad-band colors alone, and one
would have to look for some more characteristic spectral feature. It is possible
that the CB disk is significantly illuminated by the inner accretion disk,
however. This would increase its brightness, but it would also make it harder to
predict its observational appearance.

\acknowledgements

\n This research was supported in part by the National Science Foundation 
under Grant No. AST-9727875. HS acknowledges support from the European Commission
under TMR grant ERBFMRX-CT98-0195 (`Accretion onto black holes, compact objects
and prototars'). We thank the anonymous referee for detailed comments and
suggestions, in particular concerning the period gap and the effect of nova
outbursts.

\vfil\eject
\cl{\bf FIGURE CAPTIONS}
\bigskip

\figcaption{
Evolution of the surface density at the inner edge $r_{\rm i}$ of a disk 
surrrounding a mass transfering binary. The disk is fed a fraction 
$\delta=10^{-4}$ of the mass transfer rate. The
initial development up to $t_0=100$ Myr is a slow increase 
governed by the assumed constant
stellar wind angular momentum loss. After $t_0$ the 
angular momentum transfer from the binary
to the circumbinary disk causes an unstable feedback and exponential 
growth on a time scale $t_0$. 
Dotted: disk viscosity proportional to distance $r$.
Solid: same
but with viscosity increasing as $r^{1.2}$ for $r/r_{\rm i}>50$}

\figcaption{
Change of the secondary's mass with time under angular momentum loss from the
binary. 
Homologous stars are used as in Spruit and Ritter (1983). 
Solid: magnetic braking only.
Dotted: With
feedback due a circumbinary disk.}
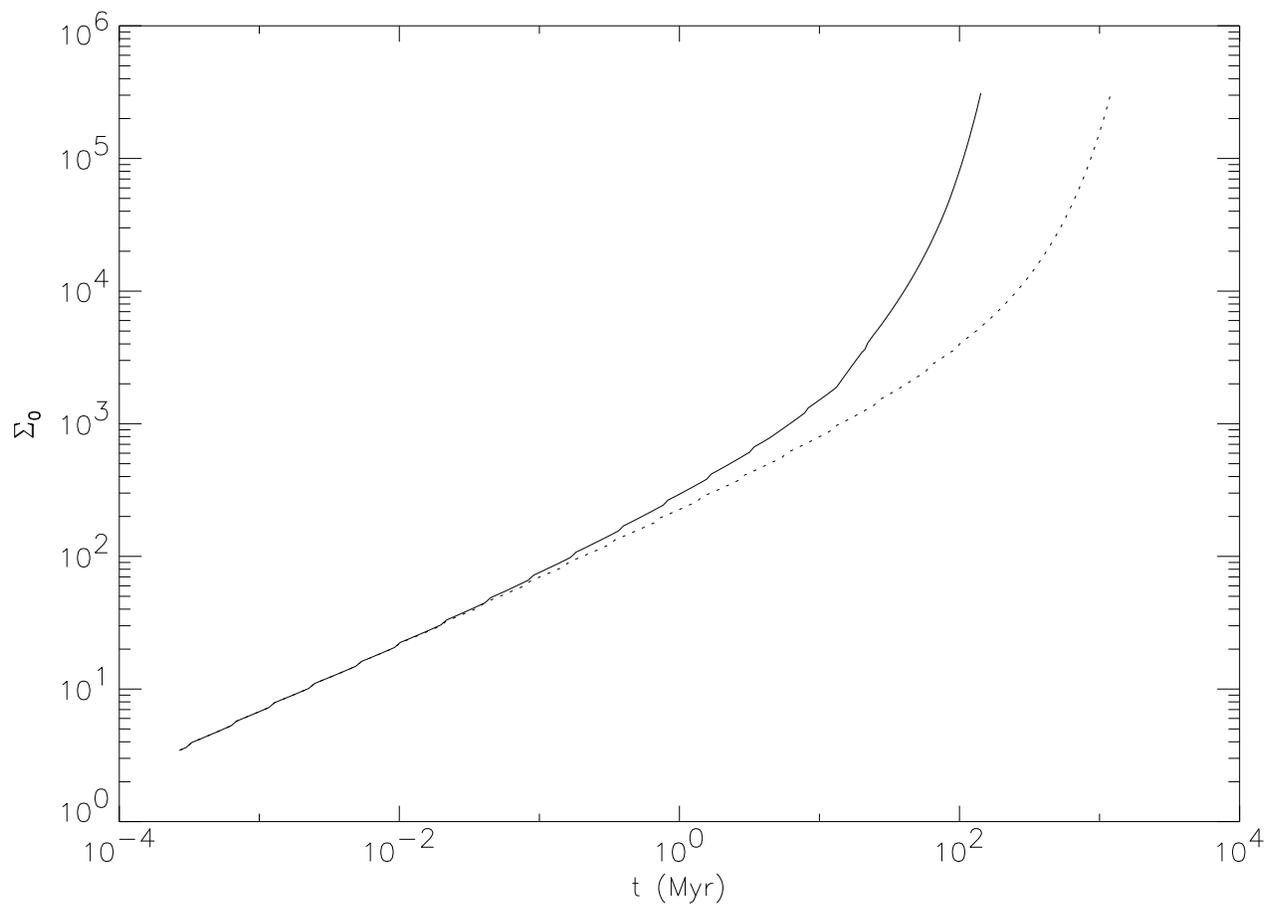
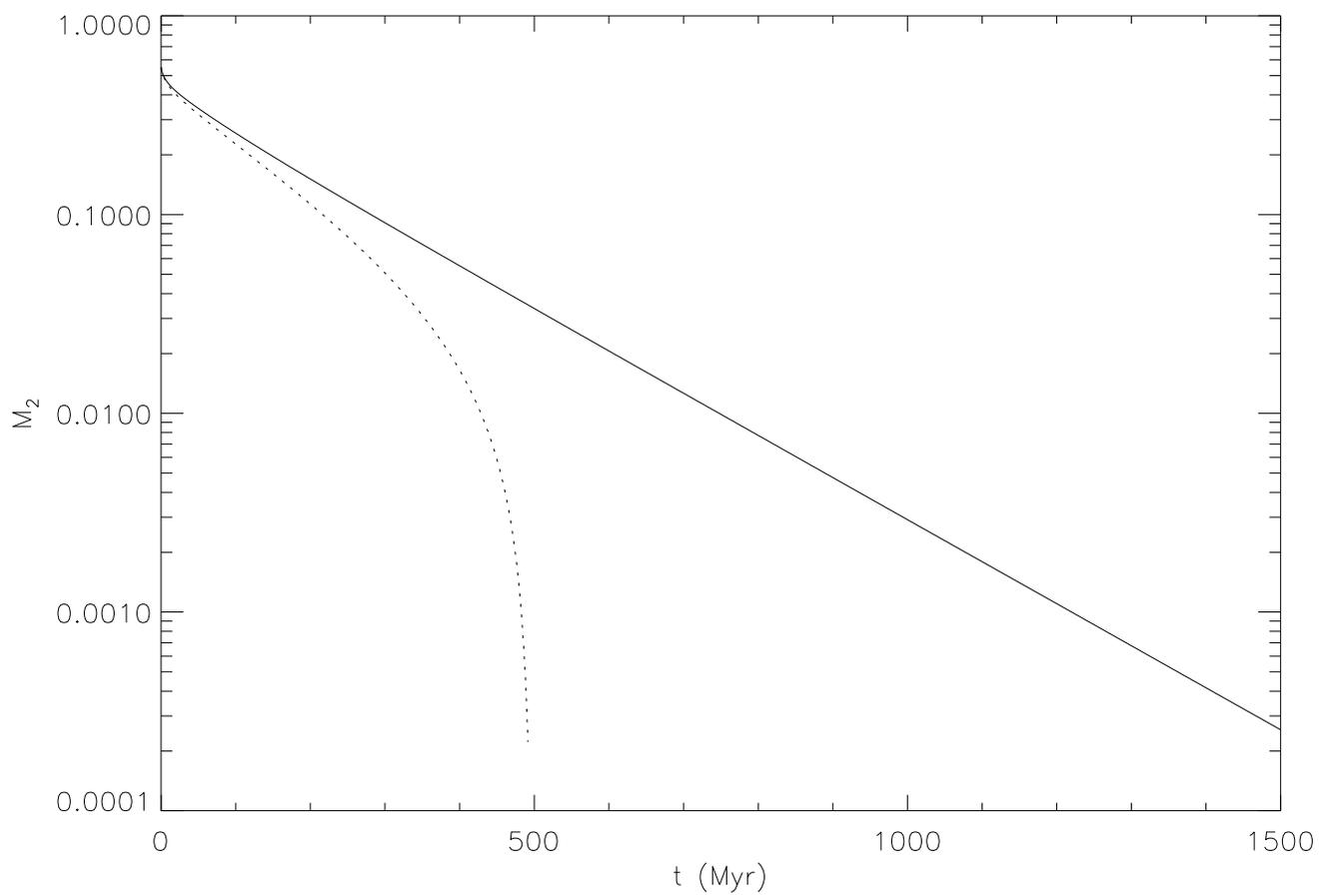
\end{document}